\documentclass[aps, showpacs,notitlepage]{revtex4-1}
\usepackage{amssymb}
\usepackage{graphics}  

\usepackage{amsfonts}
\usepackage{amsmath}
\usepackage{amssymb}
\usepackage{graphicx}
\usepackage{color}

\newcommand{\marge}[1]{\marginpar{}}  
\newcommand{\Sl}[1]{{}}           
\newcommand{\beq}[1]{\Sl{#1}\begin{equation}\if#1\empty\else\label{#1}\fi}
\newcommand{\eeq}{\end{equation}}
\newcommand{\beqa}{\begin{eqnarray}}
\newcommand{\eeqa}{\end{eqnarray}}
\newcommand{\beal}{\begin{align}}
\newcommand{\enal}{\end{align}}

\newcommand{\nm}{\nonumber\\}
\newcommand{\Eq}[1]{Eq.(\ref{#1})}

\newcommand{\la}{\langle}
\newcommand{\ra}{\rangle}

\newcommand{\Ex}[1]{(\ref{#1})}

\definecolor{red}{rgb}{1,0,0}

\providecommand{\U}[1]{\protect\rule{.1in}{.1in}}
\begin{document}
\title{Microscopic theory of anomalous diffusion based on particle interactions}
\author{James F. Lutsko and Jean Pierre Boon}
\affiliation{Physics Department, CP 231, Universit\'e Libre de Bruxelles, 1050 - Bruxelles, Belgium}
\email{jlutsko@ulb.ac.be}
\homepage{http://www.lutsko.com}
\email{jpboon@ulb.ac.be}
\homepage{ http://homepages.ulb.ac.be/~jpboon/}


\begin{abstract}
We present a Master Equation formulation based on
a Markovian random walk model  that exhibits sub-diffusion, classical diffusion and super-diffusion 
as a function of a single parameter. The non-classical diffusive behavior is generated 
by allowing for interactions between a population of walkers. At the macroscopic level, this gives rise 
to a nonlinear Fokker-Planck equation. The diffusive behavior is reflected not only in the mean-squared displacement ($\langle r^2(t)\rangle \sim t^{\gamma}$ with $0 <\gamma \leq 1.5$) 
but also in the existence of self-similar scaling solutions of the Fokker-Planck equation. 
We give a physical interpretation of sub- and super-diffusion in terms of the attractive and repulsive
interactions between the diffusing particles and we discuss analytically the limiting values of the
exponent $\gamma$. Simulations based on the Master Equation are shown to be in agreement
with the analytical solutions of the nonlinear Fokker-Planck equation in all three diffusion regimes.
\end{abstract}

\date{\today ; to appear in {\it Physical Review E}}
\maketitle

\bigskip


\bigskip

\section{Introduction}
\label{intro}

Diffusion is an ubiquitous phenomenon observed in physical, chemical, biological,
social, algorithmic systems where "objects" (particles, molecules, cells, individuals,
agents, ...) move in a seemingly random sequence of steps in such a way that their
mean squared displacement increases linearly in time: $\la r^2 \ra \sim t$ where the
proportionality factor is a constant which (apart from a numerical factor) is the diffusion 
coefficient. The resulting effect is a spread of the spatial distribution of the objects
in the form of a Gaussian whose width grows with the square root of time. 
A microscopic mechanism yielding diffusive behavior at the macroscopic scale
is based on the simple model of point particles undergoing random displacements on a 
one-dimensional lattice where they hop left or right from site to site at each tic of the clock: 
this is the random walk model for Brownian particles diffusion designed  by Einstein who
in 1905  formulated the first microscopic derivation of the diffusion equation~\cite{einstein}.

However there are many instances where the "objects" do not move freely: 
obstacles, time delays, interactions can modify their trajectories in such a way 
that the mean squared displacement deviates from the linear law and
the Gaussian structure of the dispersion is deformed or replaced by a different distribution.
So more generally,  one observes $\la r^2 \ra \sim t^{\gamma}$ where for normal diffusion ${\gamma}=1$
while if ${\gamma}\neq1$ one talks about {\em anomalous diffusion}: when ${\gamma}<1$ the 
process is said to be {\em sub-diffusive} and when ${\gamma}>1$ it is {\em super-diffusive}. 
As a result, there has been considerable interest in developing stochastic models capable of generating such behavior and the difficulty in constructing models depends on the requirements imposed. The most fundamental constraint is, of course, the need to reproduce a mean-squared displacement that exhibits power-law behavior as a function of time but this is a rather weak constraint that can be fulfilled in many ways. Perhaps the strongest demand that can be made  is for the existence of self-similar solutions which is equivalent to demanding that \emph{all} moments scale similarly, $\la r^{2n} \ra \sim t^{n\gamma}$. This implies that the distribution has the form $f(r,t) = t^{-\gamma/2} \phi(r/t^{\gamma/2})$ for some function $\phi(x)$ which is the case for classical diffusion.  Another characteristic of classical diffusion is a type of universality wherein the details of the microscopic model can be changed without affecting the macroscopic process. In the Einstein model, the jumps can be of lengths greater than one with different probabilities including  rests (jumps of length zero) without affecting the process (and, indeed, with the same diffusion constant). More generally, diverse microscopic dynamics can give rise to "diffusion" at the macroscopic level, but the underlying mechanisms may be quite different; for instance the distinction should be made between tracer motion where  experimentally one follows trajectories of distinguishable particles seeded in an active medium 
\cite{sanchez, dogariu} and the motion of tagged particles which, while identical to the medium particles, are made observable by radioactive or fluorescent markers \cite{molecularD}. The latter case is referred to as molecular diffusion which is the dynamical process considered here.

One class of diffusion models involves the use of memory or, equivalently, correlated noise since  it is easy to see that if in two successive steps, a random walker preferentially follows the first step with one in the same direction (or the opposite direction) the rate of diffusion can be dramatically altered. Non-Markovian dynamics is the mechanism behind the fractional Brownian motion \cite{fBm, reviewAnom} and certain lattice models such as the elephant random walk \cite{elephant,kumar-lindenberg}. Similarly, the use of correlated noise in the generalized Langevin equation \cite{desposito} leads to the fractional Fokker-Planck equation~\cite{FFPE_PhysToday} describing the phenomenology of anomalous diffusion in large ensembles~\cite{metzler-klafter} and for single trajectories~\cite{barkai}. 

One might wonder whether a Markovian random walk can give rise to anomalous diffusion. It is probably the case that \emph{any} type of scaling of the mean-squared displacement is possible via the application of an external force (which is to say, jump probabilities or waiting times that depend on spatial position) but the existence of a field obviously corresponds to a very particular physical circumstance. Recently, we showed that sub-diffusive behavior could also be realized in a random walk model in which the walkers interact with one-another \cite{boon-lutsko, lutsko-boon} and which is indeed Markovian and contains no external force. Starting with a rather general ansatz for the way in which interactions affect the probability of jumps we derived from the master equation, via a multiscale expansion, a nonlinear Fokker-Planck equation, a generalization of the so-called porous medium equation. When this was combined with the requirement for diffusive-like self-similar solutions, it was shown that certain limits of the jump probabilities (as explained in detail below) had to depend on the concentration of walkers in a specific power-law form. The resulting description of sub-diffusion was quite different from that given by non-Markovian models. For example, the fractional Fokker-Planck equation leads to a stretched-exponential distribution while the nonlinear model gives algebraic power-law distributions. This approach was also extended to the description of nonlinear reaction-diffusion 
systems \cite{boon-lutsko-lutsko}.

\begin{figure}
[ptb]\includegraphics[angle=-90,scale=0.3]{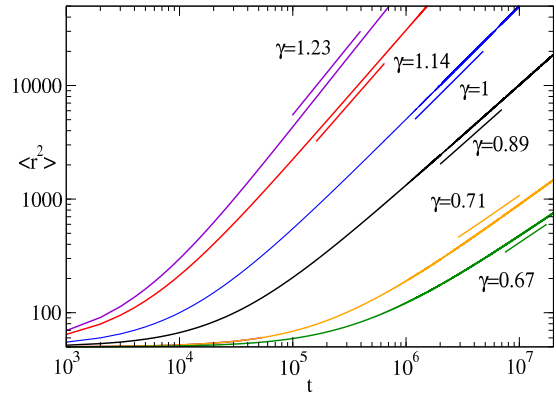}
\caption{(Color online) The mean squared displacement as a function of time obtained from the  solutions of the master equation for the cases of sub-, classical- and super-diffusion with a Gaussian initial condition. Other details of the simulation 
are described in Section \ref{simulations}. Only a single parameter varied is varied, $\alpha$ in Eq.(\ref{sol_F_alt}), and the scaling exponents are predicted from the generalized diffusion equation, Eq.(\ref{GFPE}),  to be 
$\gamma = \frac{2}{\alpha +1}$. The line segments are curves of the form $\la r^2 \ra =at^{\gamma}$ and verify the diffusive scaling with the predicted exponents at long times. In this figure the mean-squared displacement is given in units of lattice-spacings, $\delta r$, and the time in units of the fixed jump time, $\delta t$.} \label{SubSuper}
\end{figure}

In this paper, we further extend our previous work to include a description of super-diffusion. Indeed, we show that the same microscopic random walk model can exhibit sub-diffusion, classical diffusion and super-diffusion at the macroscopic scale as illustrated in Fig.\ref{SubSuper} simply by varying a single parameter.
In all cases, the behavior is ``diffusive'' in the strong sense of allowing for self-similar solutions. It will also be apparent that the model shows a kind of universality in which different microscopic dynamics yield the same macroscopic behavior. The ingredients needed to give this range of behavior are both nonlinearity and (spatial) non-locality. While our previous work showed that nonlinearity alone was sufficient to give rise to sub-diffusion, we find that spatial non-locality is critical to achieve super-diffusion within our framework. In some sense, this might be thought of as a complement or dual to the temporal non-locality of the non-Markovian approaches.

In the next Section, we review our formalism and apply it to a particular form of nonlocal interaction between walkers. In Section III, we present direct numerical solutions of the master equation and compare to the predictions of the macroscopic nonlinear Fokker-Planck equation. We show that in the super-diffusive case, considerable care is needed in extracting the continuum limit from microscopic simulations but that the two are indeed in agreement.  The paper concludes with a few comments on our results.

\bigskip

\section{Random walk model and nonlinear Fokker-Planck equation}
\subsection{The Master Equation}
\label{master}

Consider a walker moving on a one-dimensional lattice whose sites are labeled 
by integers $j=..., -2\,, -1,\,0,\,1,\,2,\,...$~. The formal expression of the microscopic 
dynamics describing the diffusive motion of Einstein's random walk model reads
\begin{equation}
n^{\ast }(r;t+1)=\xi^{(-1)} \,n^{\ast }(r+1; t)+\xi ^{(+1)}\,n^{\ast}(r-1; t)\,,  
\label{microEq}
\end{equation}%
where the Boolean variable $n^{\ast }(r;t)=\{0,1\}$ denotes the occupation
at time $t$ of the site located at position $r$ and $\xi ^{(\pm 1)}$ is a
Boolean random variable controlling the particle jump between neighboring
sites ($\xi^{(-1)}+\,\xi ^{(+1)} = 1$);  the superscript index ${(\pm1)}$ indicates
specifically that the jumps occur over one lattice distance.
Extending the possible jump steps over the whole lattice, Eq.(\ref{microEq})
becomes
\begin{equation}
n^{\ast }(r;t+1)=\sum_{j=-\infty }^{+\infty }\xi^{(j)}\,n^{\ast }(r-j; t)\,,  
\label{microEq1}
\end{equation}%
with $\sum_{j=-\infty }^{+\infty }\xi^{(j)} = 1$ (which condition prevents conflicting
occupations at the arrival site). 

The mean field description follows by ensemble averaging Eq.(\ref{microEq1})
with $\langle n^{\ast }(r;t)\rangle =f(r;t)$ and $\langle \xi ^{(j)}\rangle={P}_{j}$,  
where $j$ is the position index; using statistical independence of the 
$\xi $'s and $n^{\ast }$, we obtain 
\begin{equation}
f(r;t+\delta t)\,=\,\sum_{j=-\infty }^{+\infty }{P}_{j}(r-j\delta
r;t)\,f(r-j\delta r;t)\,,  \label{Master_Eq0}
\end{equation}%
where the distance, in lattice units, between neighboring sites is denoted by $\delta r$
and ${P}_{j}(\ell)$ is the probability of a jump of $j$ sites from site $\ell$ with $ \sum_{j}\,P_j(\ell)\,=\,1$.
Setting $j = \pm 1$ and ${P}_{\pm1} = 1/2$  in \Eq{Master_Eq0}, we have the master equation 
for the usual random walk wherefrom Einstein derived the classical diffusion equation \cite{einstein}.
Note that \Eq{Master_Eq0} can also be rewritten as
\beq{me2}
f(r, t + \delta t) - f(r, t) = \sum_{j }\,\left( P_j(r - j\,\delta r)\,f(r - j\delta r, t) -  P_j(r) \,f(r, t) \right) \,,
\eeq
which expresses the rate of change of the particle distribution as the difference
between the incoming  and  outgoing fluxes at location $r$. 
Equation (\ref{Master_Eq0}) generalizes Einstein's master equation providing a formulation 
which in the hydrodynamic limit leads to the description of non-classical diffusion:
when  $P_j(\ell)$ has a functional form ${p}_j\,{F}[f (r, t)] \sim {p}_j\,f^{\alpha - 1}$ (where ${p}_j$ is a 
prescribed spatial distribution), \Eq{Master_Eq0} becomes the master equation leading to the description 
of {\em nonlinear diffusion}~\cite{boon-lutsko}.

\subsection{The Fokker-Planck equation}
\label{fokker-planck}

We now introduce the further generalization that the jump probabilities 
$P_j(\ell)$  take a functional form depending on both concentrations 
at the starting point and at the end point of the jump
\beq{P(ff)}
P_j = {p}_j\,{F}[f(r - j\,\delta r, t), f (r, t)]\,,\;\;\;\;\;\mbox{with}\;\;\;\;\sum_{j}\,\,P_j = 1\,,
\eeq
where the probabilities ${p}_j$ are drawn from a prescribed distribution and the bounding 
condition $0\leq P_j \leq1$ imposes $0\leq  F(x,y) \leq1$ as well as restrictions on the 
functional form of $F(x,y)$. Under these conditions, multiscale expansion of the master 
equation was shown to give the generalized Fokker-Planck (or generalized diffusion) equation \cite{lutsko-boon}
\beqa
\frac{\partial f}{\partial t}+M_1\frac{\partial}{\partial r}\left[  xF\left(
x,x\right)  \right]  _{f}  &=& M_2\frac{\partial}{\partial r}\left[
\frac{\partial xF\left(  x,y\right)  }{\partial x}-\frac{\partial xF\left(
x,y\right)  }{\partial y}\right]  _{f}\frac{\partial f}{\partial
r}\nonumber\label{main2}\nm
&+&\frac{1}{2}M_1^{2}\delta t\frac{\partial}{\partial r}\left[  \frac{\partial
xF\left(  x,y\right)  }{\partial x}-\frac{\partial xF\left(  x,y\right)
}{\partial y}-\left(  \frac{\partial xF\left(  x,x\right)  }{\partial x}%
\right)  ^{2}\right]  _{f}\frac{\partial f}{\partial r}\,,
\label{GFPE}
\eeqa
with the notation 
\begin{equation}
\left[  \frac{\partial xF\left(  x,y\right)  }{\partial x}\right]
_{f}=\left[  \frac{\partial xF\left(  x,y\right)  }{\partial x}\right]
_{x=f\left(  r,t\right)  ,y=f\left(  r,t\right)  }\,.
\end{equation}
{{In \Eq{GFPE}, $M_1$ and $M_2$ are given by }}
\begin{align}
&M_1  = \frac{\delta r}{\delta t}\,\sum_{j}\,j\,p_{j} = \frac{\delta r}{\delta t}\,J_{1}\,,\nonumber\\
&M_2=\frac{1}{2}\frac{\left(  \delta r\right)^{2}}{\delta t}\left( \left(\sum_{j}\,j^2\,p_{j}\right) - %
J_{1}^{2}\right)  = \frac{1}{2}\frac{\left(  \delta r\right)^{2}}{\delta t}\left(  J_{2}-J_{1}^{2}\right) 
\end{align}
{{where the $J_{n}$'s denote the moments $J_{n}=\sum_{j}j^{n}p_{j}$. }}
Note that for $F\left( x,y\right) = 1$, \Eq{GFPE} reduces to the classical 
advection-diffusion equation.

Since the function $F\left(  x,y\right)$ is defined in terms of the jump probabilities, 
it must be bounded, and  so must satisfy 
\begin{equation}
0\leq xF\left(  x,y\right)  \leq1\;\;\;\;\mbox{and}\;\;\;\;0\leq yF\left(
x,y\right)  \leq1\,,\;\;\;\;\forall \;x,y\in\left[  0,1\right]\,. 
\label{bounds}
\end{equation}
It was shown in \cite{lutsko-boon} that self-similar solutions are possible if and only if 
\begin{equation}
\lim_{y \rightarrow x} \left[\frac{\partial xF\left(  x,y\right)  }{\partial x}-\frac{\partial xF\left(
x,y\right)  }{\partial y}\right] \sim x^{\alpha-1}
\end{equation}
where the scaling exponent $\alpha$ is related to the diffusion exponent by
\begin{equation}
\gamma=\frac{2}{\alpha + 1}\;.
\end{equation}
Our previous work focused on \emph{local} models for which $F(x,y) \equiv F(x)$ in which case 
the bounds given above \Ex{bounds} demand that
$\alpha \geq 1$ and therefore $0\leq\gamma\leq1$, which is the correct  
formulation of sub-diffusion \cite{RDapplication} but excludes the case of super-diffusion
 ($\gamma > 1\;\;\longleftrightarrow \;\;\alpha < 1$). 
 
 \section{Sub-diffusion and super-diffusion}
 \subsection{A model  for sub- and super-diffusion}
 \label{FP_sub_super}

Both types of anomalous diffusion (sub- and super-) can be described in a single 
formulation when the jump probabilities depend on the occupation probabilities 
at both the starting point and the end point of the jump by means of the ansatz
\begin{equation}
{F}(x,y; \omega_s, \omega_e) \sim \omega_s\,F_s(x) + \omega_e \,F_e(y) \,,
\end{equation}
where $\omega_s$ and $\omega_e$ are weighting factors relative to the functionals 
of the concentrations at the starting point and at the end point of the jump. Since an overall scale factor is unimportant,
we can divide by  $\omega_s$  and write $a \equiv \omega_e/\omega_s$ without loss of generality.
Finally, in order to assure that the constraint (\ref{bounds}) is satisfied, we normalize to give the ansatz
\begin{equation}
F(x,y;a) = \frac{F(x) + aF(y)}{F(x)+F(y)}.
\label{model_alt}
\end{equation}
The positivity arguments, $x,y$ and the constraints (\ref{bounds}) and $F(x,y;a) \ge 0$ imply
 that $0 \le a \le 1$. We emphasize that this ansatz is motivated by simplicity and the necessary mathematical requirements.

Considering the case that there is no drift ($M_1=0$ in \Eq{GFPE}), and in order that  the general 
formulation describe diffusion, we should have a scaling solution of the form
$f\left(  r,t\right)  =t^{-\gamma/2}\phi\left(  r/t^{\gamma/2}\right)$;
it was shown in  \cite{lutsko-boon} that the scaling hypothesis demands that
\begin{equation}
\lim_{y\rightarrow x}\left( \frac{\partial }{\partial x}xF\left( x,y; a \right) -%
\frac{\partial }{\partial y}xF\left( x,y; a, \alpha\right) \right) = K\, x^{\alpha - 1 } \,,
\label{scaling_cond}
\end{equation}%
for some constant $\alpha$. Using the model functional \Ex{model_alt} in the l.h.s of \Ex{scaling_cond} gives
\begin{align}
\frac{1 + a}{2} + \frac{1-a}{2} \frac{x\,F^{\prime }\left( x\;a,\alpha \right) }{\,F\left(x;a,\alpha\right) }& = K\,  x^{\alpha - 1 }  \,,
\end{align}%
which is solved to yield
\begin{align}
F\left( x;a,\alpha\right)& =\frac{B}{x^\frac{1+a}{1 -a}}\,\exp {\left( \frac{2\,K  }%
{1-a}\,\frac{x^{\alpha - 1 }}{\alpha - 1}\right) } \,,
\label{fx} 
\end{align}%
where $B$ is an integration constant; reinserting \Ex{fx}  into  \Ex{model_alt}, we find
\begin{align}
F(x,y; a, \alpha) = \frac{1 + a \,\left(\frac{x}{y}\right)^{\frac{1+a}{1-a}}\,\exp \left( \frac{2\,K} %
{1-a}\,\frac{y^{{\alpha -1}}-x^{{\alpha -1}}}{\alpha -1}\right) }%
{1+  \left(\frac{x}{y}\right)^{\frac{1+a}{1-a}}\,\exp \left( \frac{2\,\lambda  } %
{1-a}\,\frac{y^{{\alpha -1}}-x^{{\alpha -1}}}{\alpha -1}\right)} \,.
\label{sol_a}
\end{align}%

The natural limit: $\lim_{{{\alpha \rightarrow 1}}}F(x,y;a)=1$ requires { $ K= \frac{1}{2}(1 + a)\, \lambda  ^{\alpha-1}$ where $\lambda  $ is an arbitrary constant with units of length.  Thus,}
\begin{align}
F(x,y; a, \alpha) = \frac{1 + a\, G(x,y; a, \alpha)}{1 +  G(x,y; a, \alpha)} \,,
\label{F_G}
\end{align}%
with{
\begin{align} 
{G(x,y; a, \alpha)} \,=\,
\left(\frac{x}{y}\right)^{\frac{1+a}{1-a}}\,\exp \left( \frac{1+a}{1-a}\; %
\lambda^{\alpha-1} \,\frac{y^{{\alpha -1}}-x^{{\alpha -1} }}{\alpha -1} \right)\;.
\label{sol_F_alt}
\end{align}%
}
It is clear that the limits $a=1$ and $\alpha \rightarrow 1$ give normal diffusion
($F(x,y;a, \alpha)=1$) and, for any finite value of $x>0$, we have
$0\leq F(x,y;a, \alpha)<1 \;\;, \forall \; y \in \left[0,1\right]\,.$
Furthermore, one has that%
\begin{equation} \label{zerox}
\lim_{x\rightarrow 0}F(x,y; a, \alpha)=\left\{ 
\begin{array}{c}
1,\;\;\;\alpha > 1 \\ 
a,\;\;\;\alpha < 1%
\end{array}%
\right.
\;\;\;\;\;\mbox{and}\;\;\;\;%
\lim_{y\rightarrow 0}F(x,y; a, \alpha)=\left\{ 
\begin{array}{c}
a,\;\;\;\alpha > 1 \\ 
1,\;\;\;\alpha < 1  %
\end{array} 
\right.
\end{equation}

To provide some interpretation for this model, we note that%
{
\begin{align}
\frac{\partial }{\partial x}F(x,y; a, \alpha)  
& = (1+a)\, \frac{G(x,y; a, \alpha)}{\left({1 +  G(x,y; a, \alpha)}\right)^2}\,\left( \frac {(\lambda \, x)^{\alpha - 1} - 1}{x}\right)\;,
\end{align}
\begin{align}
\frac{\partial }{\partial y}F(x,y; a, \alpha)   
& = (1+a)\, \frac{G(x,y; a, \alpha)}{\left({1 +  G(x,y; a, \alpha)}\right)^2}\,\left( \frac {1-(\lambda \, y)^{\alpha - 1}}{y}\right)\;.
\end{align}
}
Since $0<x,y<1$, the signs of these derivatives are determined by the factors on the right:{
$\left( (\lambda\,  x)^{\alpha - 1} - 1\right)$ and $\left( 1 - (\lambda \, y)^{\alpha-1} \right) $}, and so depend on whether 
($\alpha -1$)  is positive or negative:
\begin{align}
\alpha & > 1\Longrightarrow \frac{\partial }{\partial x}F(x,y; a, \alpha)<0<\frac{\partial 
}{\partial y}F(x,y; a)  \notag \\
\alpha & < 1\Longrightarrow \frac{\partial }{\partial x}F(x,y; a, \alpha)>0>\frac{\partial 
}{\partial y}F(x,y; a)  
\end{align}%

In the first case, $\alpha  > 1$, the jump probability decreases with the concentration
at the starting point and increases with the concentration at the arrival site; in other
words the jump rate is reduced by putting more walkers at the origin and increased 
by putting more at the terminus of a jump: this is analogous to an attractive interaction. 
For $\alpha < 1$, we have the reverse situation: the jump rate is increased by putting 
more walkers at the origin and decreased by putting more at the terminus, thus 
emulating a repulsive interaction. In the standard problem with all walkers at the origin 
at $t=0$, the distribution decays monotonically away from the origin; thus, if the particles 
repel, the distribution expands faster (i.e. tends to a uniform distribution more
quickly) whereas if they attract, then this attraction slows down the spread of the distribution.
The physical interpretation is that  attractive interactions give sub-diffusion and 
repulsive interactions give super-diffusion.

\subsection{Explicit form of the scaling solutions}

{{In the absence of drift, the generalized advection-diffusion equation, Eq.(\ref{GFPE}) with \Ex{F_G} 
and \Ex{sol_F_alt}, reads
\begin{equation}
\frac{\partial f}{\partial t}=%
\lambda  ^{\alpha-1} \frac{1+a}{2\,\alpha}\,M_2 \frac{\partial^2 }{%
\partial r^2}\,f^{\alpha} \,,
\label{GADE1}
\end{equation}%
(here $M_2 = \frac{1}{2}\frac{\left(  \delta r\right)^{2}}{\delta t}\,\sum_{j}\,j^2\,p_{j}$)
and the scaling solutions are obtained following the development given in ref.\cite{lutsko-boon}
yielding
\begin{equation}
f\left( r, t\right) =t^{-\gamma /2}\,W\left( 1\pm V\,\frac{r^{2}}{t^{\gamma }}\right) ^{1/(\alpha-1)}\,,
\label{sol_GADE1}
\end{equation}%
 where $W$ and $V$ are be determined by the normalization condition (see below) and the expression 
 obtained  by inserting \Ex{sol_GADE1} into \Ex{GADE1}
\begin{equation}
V W^{\alpha -1} = \lambda  ^{1-\alpha}\,\frac{|1-\alpha |}{1+\alpha}\,\frac{1}{1+a}\,{M_2}^{-1}\,. 
\label{VW1}
\end{equation}%

 \noindent(i)  For the super-diffusive case, $\alpha <1$, %
\begin{equation}
f\left( r, t\right) =t^{-\gamma /2}\,W\left( 1+V\,\frac{r^{2}}{t^{\gamma }}\right) ^{1/(\alpha-1)}\,.
\end{equation}
The normalization condition (using the reduced variable $\zeta = V^{1/2}\,\frac{r}{t^{\gamma/2}}$) reads
\begin{equation*}
W\,V^{ -1/2}\,\int_{-\infty}^{\infty }\,d\zeta\,\left( 1+\zeta^{2}\right) ^{1/(\alpha-1) } =1  \;\;\;\Longrightarrow\;\;\;
\frac{W}{\sqrt V} = \frac {1}{\sqrt \pi} \, \frac{\Gamma(\frac{1}{1-\alpha})}{\Gamma(\frac{1}{1-\alpha} - \frac{1}{2})}
\end{equation*}%
 provided 
\begin{equation}
\frac{\alpha +1}{\alpha-1 }<0\;\;\;\Longrightarrow\;\;\; \alpha > -1\, \;\;\;\Longrightarrow\;\;\; 
\gamma > 1 \,, %
\end{equation}%
and the mean-squared displacement 
$\left\langle r^{2}\right\rangle  =\int_{-\infty }^{\infty }r^{2}f\left(r;t\right) dr$ is given by
\begin{equation}
\left\langle r^{2}\right\rangle =t^{\gamma }\,W\,V^{-3/2} %
\int_{-\infty }^{\infty } d\zeta \, \zeta^{2}\left(1+\zeta^{2}\right) ^{1/(\alpha-1) } \,  \;\;\Longrightarrow\;\;
\left\langle r^{2}\right\rangle = \,\frac{W}{\sqrt V} \,\frac{t^{\gamma }}{V}\,\frac{\sqrt \pi}{2}%
\frac{\Gamma(\frac{3\alpha - 2}{2(1-\alpha}))}{\Gamma(\frac{1}{1-\alpha})} %
\,=\,\frac{t^{\gamma }}{V} %
 \, \frac{\Gamma(\frac{3\alpha - 2}{2(1-\alpha}))}{2\,\Gamma(\frac{1+\alpha}{2(1-\alpha}))} \,,
\label{msd_super}
\end{equation}%
which is finite if 
\begin{equation} \label{bound}
\frac{3\alpha - 1}{\alpha-1 }<0\;\;\;\Longrightarrow\;\;\; \alpha >\frac{1}{3}\;\;\;\Longrightarrow\;\;\; \gamma <\frac{3}{2} \,.
\end{equation}

 \noindent(ii)  for the sub-diffusive case, $\alpha >1$,  the distribution has finite support so that 
\begin{equation} \label{dist1}
f\left( r, t\right) =t^{-\gamma /2} \,W\left( 1- V\,\frac{r^{2}}{t^{\gamma }%
}\right) ^{1/(\alpha-1) }\Theta \left( 1- V\,\frac{r^{2}}{t^{\gamma }}\right) \,, 
\end{equation}%
with the normalization condition%
\begin{equation*} 
W\,V^{ -1/2}\,\int_{-1}^{1}\,d\zeta\,\left( 1 - \zeta^{2}\right) ^{1/(\alpha-1) } =1 \;\;\;\Longrightarrow\;\;\;
\frac{W}{\sqrt V} = \frac {1}{\sqrt \pi} \, \frac{\Gamma (\frac{2\alpha}{\alpha -1}  + 1)}{\Gamma (\frac{1}{\alpha -1} + 1)}
\end{equation*}%
and the mean-squared displacement %
\begin{equation}
\left\langle r^{2}\right\rangle =t^{\gamma }\,W\,V^{-3/2} %
\int_{-1}^{1} d\zeta \zeta^{2}\left(1- \zeta^{2}\right) ^{1/(\alpha-1) } \;\;\Longrightarrow\;\;
\left\langle r^{2}\right\rangle = \,\frac{W}{\sqrt V} \,\frac{t^{\gamma }}{V}\, \frac{\sqrt \pi}{2}\,
\frac{\Gamma(\frac{1}{\alpha -1} +1)}{\Gamma(\frac{1}{\alpha-1} +\frac{5}{2})} %
\,=\,\frac{t^{\gamma }}{V} \,
 \, \frac{\Gamma(\frac{3\alpha -1}{\alpha -1})}{2\,\Gamma(\frac{5\,\alpha - 3}{\alpha -1} )} \,.
\label{msd_sub}
\end{equation}%
The continuum results therefore imply that an initial distribution of the form
$W^{\prime}\,(1+s_{\alpha }\left( r/w\right) ^{2})^{1/(\alpha-1) }\Theta \left( 1+s_{\alpha
}\left( r/w\right) ^{2}\right) ,$ with $W^{\prime}$ determined by normalization and $s_{\alpha} = \mp$ for $\alpha \gtrless 1$, will evolve self-similarly with mean-squared displacement increasing as $t^{\gamma }$. 
Indeed for both sub- and super-diffusion, the mean-squared displacement takes the form
\begin{equation}
\left\langle r^{2}\right\rangle = \widetilde{D}_{\gamma} \, t^{\gamma } \,,
\label{msd}
\end{equation}
with
\begin{equation}
 \widetilde{D}_{\gamma} = {\mbox{const}}\times  W\,V^{-3/2} = 
{\mbox{const}} \times \lambda  ^{2(1- \gamma )}\,\left(%
\frac{1+a}{ \gamma - 1} \right)^{\gamma}\,{M_2}^{\gamma}\,.
\end{equation}
$\widetilde{D}_{\gamma}$ is the {\em anomalous  diffusion coefficient} and has dimensions 
$[\widetilde{D}_{\gamma}] = L^2 \, T^{-\gamma}$.  

Determining the quantities $W$ and $V$ from the normalization condition combined with \Ex{VW1}, 
the distribution function can be written more explicitly as
\begin{equation}
f\left( r, t\right) = \,\frac{1}{\sqrt {m_{\alpha}\,\pi {\widetilde{D}_{\gamma}\,{t^{\gamma}}}}}
\left( 1 + \,\frac{r^{2}}{m_{\alpha}\,\widetilde{D}_{\gamma}\,t^{\gamma }}\right) ^{1/(\alpha-1)}
\;\;\;;\;\;\;\; {\alpha} < 1 \;,
\label{sol_GADEsuper}
\end{equation}%
for super-diffusion,  and
\begin{equation}
f\left( r, t\right) = \,\frac{1}{\sqrt {n_{\alpha}\,\pi {\widetilde{D}_{\gamma}\,{t^{\gamma}}}}}
\left( 1- \,\frac{r^{2}}{n_{\alpha}\,\widetilde{D}_{\gamma}\,t^{\gamma }}\right) ^{1/(\alpha-1)}\,
\Theta \left( 1-  \,\frac{r^{2}}{n_{\alpha}\,\widetilde{D}_{\gamma}\,t^{\gamma }}\right) 
\;\;\;;\;\;\;\; {\alpha} > 1 \;,
\label{sol_GADEsb}
\end{equation}%
for sub-diffusion  ($ m_{\alpha}$ and  $n_{\alpha}$ are constants). 

Similarly \Eq{GADE1} can be written as
\begin{equation}
\frac{\partial }{\partial t} f(r, t)= \, \frac{\partial }{\partial r} {\cal{J}}_{\alpha}(r, t)\,%
 \;\;\;\;\;\mbox{with} \;\;\;\;\; {\cal{J}}_{\alpha}(r, t)\,=\,%
{\cal{D}}_{\alpha}\,\frac{\partial }{\partial r} f(r, t) \,,
\label{GADE3}
\end{equation}%
where ${\cal{J}}_{\alpha}(r, t)$ is the current density and
${\cal{D}}_{\alpha} = \frac{1+a}{2}\,\left(\lambda  \,f \right)^{\alpha-1} M_2$ 
has the dimensions of a diffusion coefficient ($ L^{2} \, T^{-1}$). 
It follows that we have the relation 
$ \widetilde{D}_{\gamma}\,=\,c_{\gamma}\,{\cal{D}}_{\alpha}^{\gamma}$,
where $c_{\gamma}$ is a constant with $\lim_{\gamma \rightarrow 1} \, c_{\gamma}\, =1$.
These results emphasize that the anomalous diffusion coefficient $\widetilde{D}_{\gamma}$
cannot be defined in the usual sense $\lim_{t\rightarrow \infty} \frac{\la r^{2}(t)\ra}{t} = D$ 
(which would give the unphysical values $D=0$ for sub-diffusion and $D=\infty$ for super-diffusion), 
but should be considered in terms of the mean squared displacement 
\Ex{msd} obtained from the probability distribution function  or by simulation of the master equation
giving a physically observable quantity as shown in Fig.1.
Only in the limit ${\gamma = \alpha =1}$  and $a=1$ does one have the  the classical result:
$\widetilde{D}_{\gamma =1}\,=\,{\cal{D}}_{\alpha =1}\,=\,M_2 $  
 with dimension $ L^2 \, T^{-1}$ \cite{NoteC}; in this limit \Eq{GADE1} is the usual diffusion equation
and $f(r, t)$ takes the classical Gaussian form $f\left( r, t\right) =  \left(4\,\pi {D}\,{t} \right)^{-1/2}
\exp \left(- \,\frac{r^{2}}{4\,{D}\,t}\right)$ with $D = {\cal{D}}_{\alpha =1}$.

\section{Numerical simulations}
\label{simulations}

\subsection{Comparison between the master equation and the nonlinear Fokker-Planck equation}

Because the macroscopic description (the nonlinear Fokker-Planck equation) is derived from the microscopic dynamics (i.e. the random walk model) by means of a well-defined but approximate multiscale expansion it is interesting to compare the two to verify that they are indeed in agreement for the new ansatz proposed here. We therefore wish to compare the
result of the direct, numeric, solution of the master equation to the
continuum model, the generalized diffusion equation, {{Eq.(\ref{GADE1})}}, which predicts the
existence of exact scaling solutions (the explicit forms of which are given in section III.B).

In the following, we use the elementary probabilities $p_{\pm i}=p$ for $%
1\leq i\leq n$ for some values of $p$ and $n$ and have set $a=0$. 
Following our previous work \cite{lutsko-boon}, we will use $n=2$ (although in fact we have obtained similar results for one-step jumps, $n=1$). We
found that the simulations were increasingly sensitive to the value of $p$ as $\alpha$ 
decreased: in fact, when this parameter is too large, and when $\alpha <1$ (super-diffusive case) 
the distribution determined from the master equation did not
decay smoothly but, rather, showed an oscillatory structure in both space
and time with a spatial period of several lattice sites. This is because in this
case the walkers repel each other. So, if in an initial configuration the
probability satisfies $f_{n}>f_{n+1}$ then so much more probability can get
transferred from site $n$ to site $n+1$ than is transferred from $n+1$ to $n+2
$ that one finds $f_{n+1}>f_{n}$. In the next time step, this causes a flow
in the opposite direction. We therefore fixed on a value of $p=10^{-4}$
which is small enough to avoid any lattice-scale oscillatory behavior for $\alpha \ge 0.5$ 
although we could use, e.g. $p=10^{-3}$ for $\alpha \ge 0.75$.

Even with this problem under control, we did not immediately obtain
agreement between the lattice and continuum models. The left panel of Fig.\ref{distributions} 
shows several snapshots of the distribution for the case $\alpha =0.5$ $(\gamma =4/3)$
starting with a distribution of width $w=10$. While the two distributions
are similar, there are significant differences between them. This was
reflected in the mean-squared displacement from a run of $1$ million updates
of the master equation which appeared to behave as a power-law but with an
exponent $\gamma$ of approximately $1.5$ which is considerably larger than expected.
Trying a wider initial distribution improved the agreement with the
theoretical distribution (see Fig.\ref{distributions}; right panel)\ but still gave an exponent of about $%
1.45$ for the mean-squared displacement. In order to test whether this is an
equilibration effect, we used runs of $10$ million time steps and extracted
the mean squared displacement from each window of $1$ million steps:\ the
results are shown in Fig.\ref{gamma_t}. It is clear that several million time steps are
required for the exponent of the mean-squared displacement to stabilize but
even then, the final value is significantly above the theoretical value.
Finally, in Fig.\ref{gamma_w}, we show the long-time value (from the final window of $1$
million updates) as a function of the initial width of the distribution.
Fitting to a function of the form $\gamma \left( w\right) =\gamma _{\infty
}+a\left( \frac{1}{W}\right) ^{b}$ gives for $\alpha =0.5$ a value of $\gamma
_{\infty }=1.34$ and for $\alpha =0.65$ a value of $\gamma _{\infty }=1.23$
which compare well with the theoretical values of $1.33$ and $1.21$
respectively.  

\begin{figure}
[ptb]\includegraphics[angle=-90,scale=0.3]{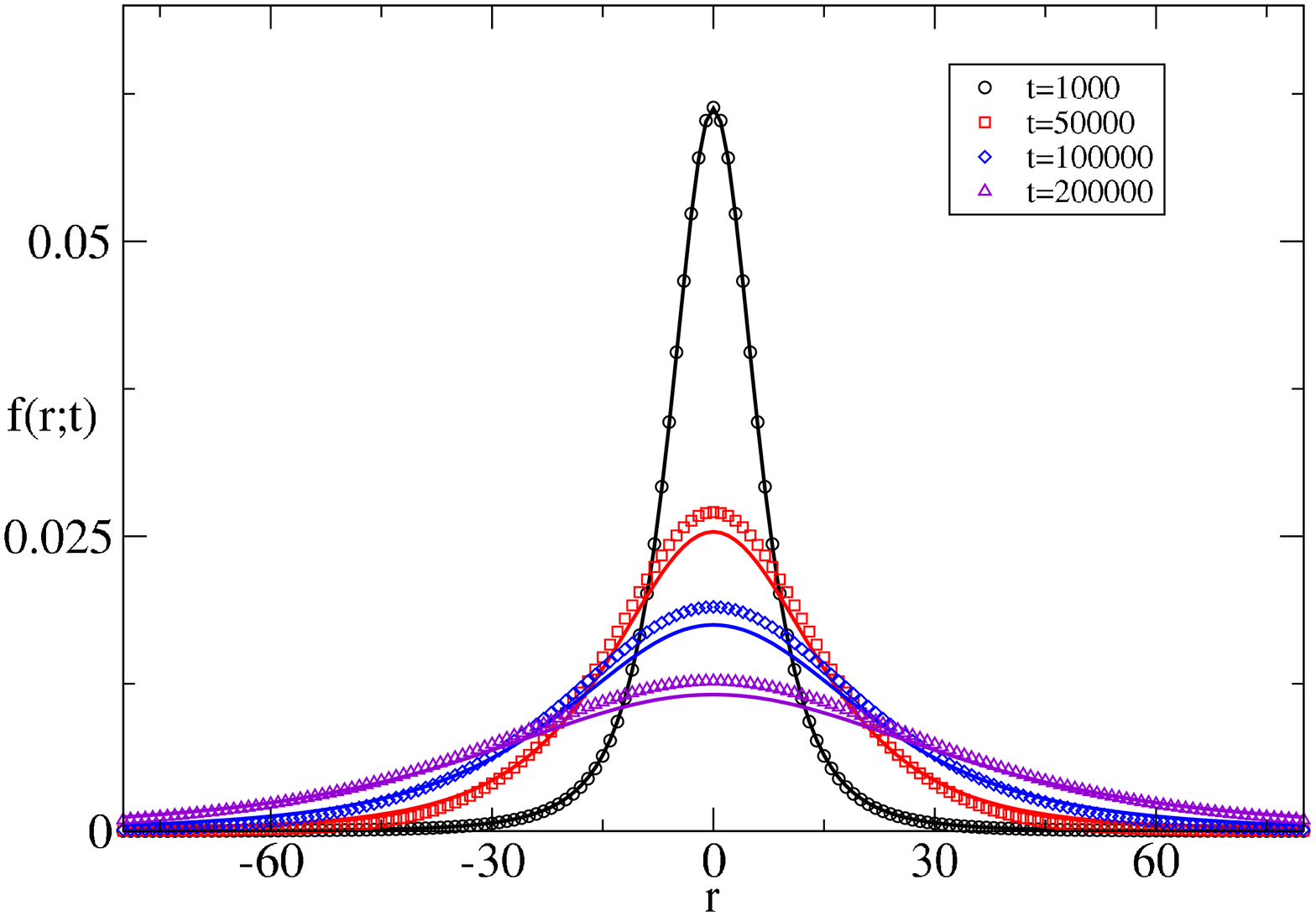}
\includegraphics[angle=-90,scale=0.3]{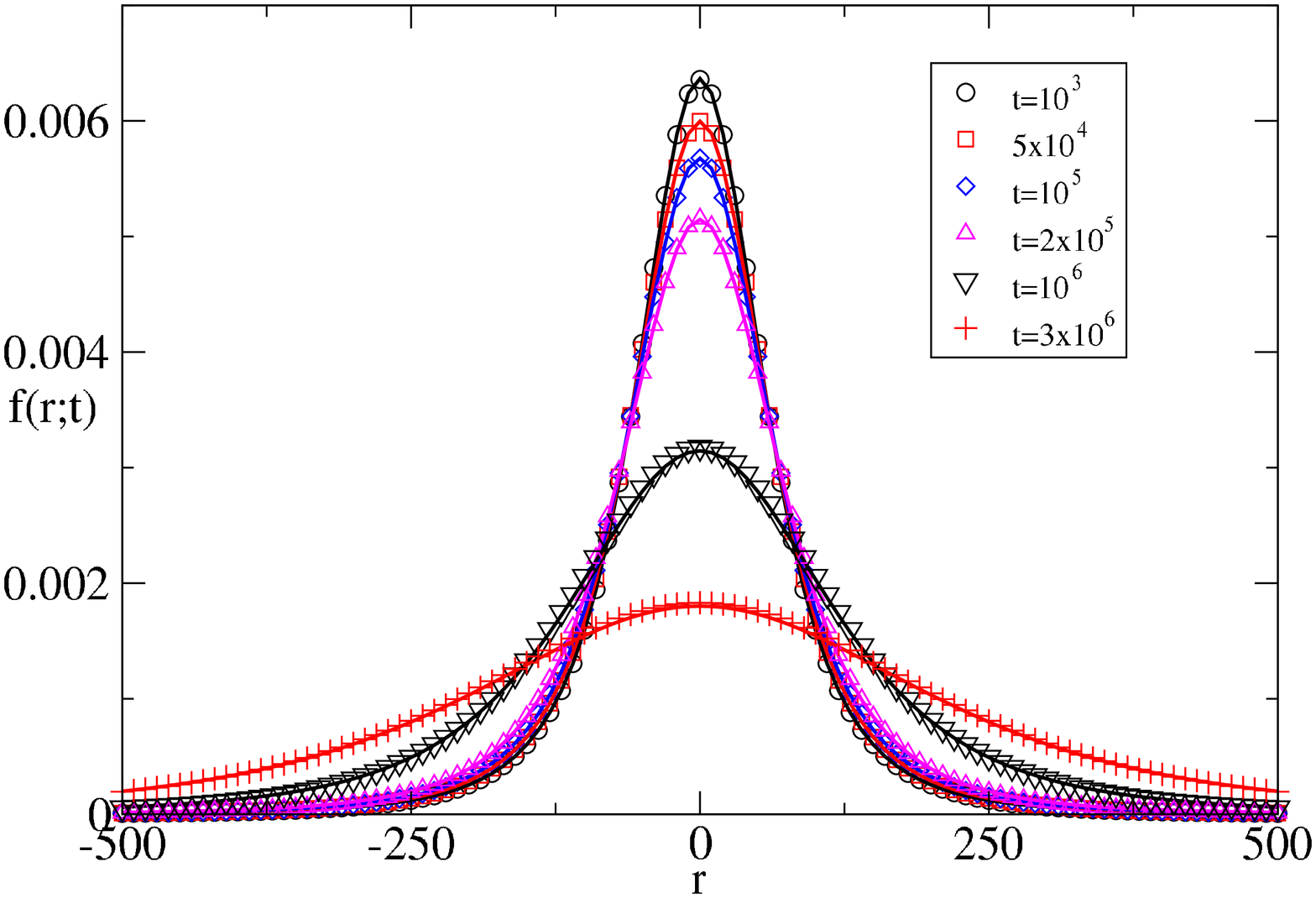}
\caption{(Color online) The spatial distribution as a function of time as determined by direct numerical solution of the master equation (symbols) and the analytic distribution, Eq.(\ref{dist1}) for the case $\alpha = 0.5$ corresponding to a scaling exponent of $\gamma = 4/3$. All quantities are dimensionless (i.e. $\delta r = \delta t = 1$). In both cases, the initial distribution was the predicted scaling solution as described in the text: for the figure on the left, the initial width was $w=10$ while for the figure on the right it was $w=100$.} 
\label{distributions}
\end{figure}

Why should the result be so sensitive to the width of the distribution?
Note that in order to obtain the continuum limit, we have to expand the
distribution $f\left( x+n\right) $ in terms of $f\left( x\right) $. However,
the form of the jump probability actually requires that we evaluate terms of
the form%
\begin{equation*}
f^{(\alpha-1) }\left( x+n\right) -f^{(\alpha-1) }\left( x\right) =(\alpha-1) f^{\alpha-2}\left(
x\right) f^{\prime }\left( x\right) n+...
\end{equation*}%
The point is that with $\alpha <1,$ and given that $f(x)<1$,  the coefficient $%
f^{\alpha-2}\left( x\right) $ can be very large so that the expansion is only
reasonable if $f^{\prime }(x)$ is sufficiently small which, in turn, requires a wide 
enough distribution in units of lattice spacing. 

In summary, to get agreement with the continuum limit for the smallest values of $\alpha$ considered here - which is to say, for the strongest super-diffusivity - it is necessary to begin with very wide distributions and to perform sufficiently long runs. However, it is important to emphasize that these statements are quantified in terms of lattice spacings and time-steps: this has nothing to do with the \emph{physical} width of the distribution and the \emph{physical} duration of the process since both of these depend on the, as yet unspecified, $\delta r$ and $\delta t$. As the distribution becomes wider and the times become longer, we can decrease both of these quantities so that the physical width of the distribution and time of the process remain constant (and as well  maintaining the ratio $\frac{\delta r^2}{\delta t}$ which determines the physical diffusion constant). Hence, the issues explored in this section pertain to reaching the continuum limit and do not indicate any constraints on the physics of the model. 

\begin{figure}
[ptb]\includegraphics[angle=-90,scale=0.3]{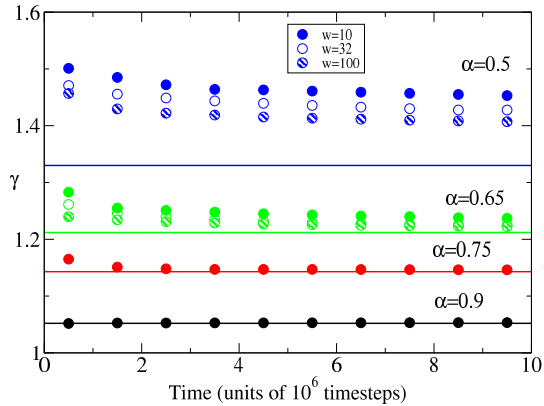}
\caption{(Color online) The scaling exponent, $\gamma$, as determined from fits of the numerical solution of the master equation. The fits were to the predicted functional form $<r^2>= a_{0}(a_{1}+t)^{\gamma}$ and were performed independently with successive windows of $10^{6}$ timesteps. The figure also shows results for different initial widths, $w$, (in units of lattice spacings) as described in the text. The lines are drawn at the predicted values $\gamma = \frac{2}{1+\alpha}$.} \label{gamma_t}
\end{figure}

\begin{figure}
[ptb]\includegraphics[angle=-90,scale=0.3]{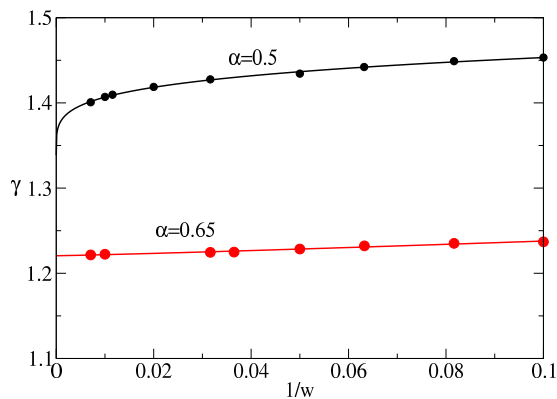}
\caption{(Color online) Gamma as a function of the inverse of the initial width (in lattice units). The lines are fits to $\gamma(w) = a_{0}+a_{1}\left( \frac{1}{w} \right)^{a_{2}}$. For $\alpha=0.5$, the fit gives $\gamma(w\rightarrow \infty) = 1.339$ and for $\alpha=0.65$, $\gamma(w \rightarrow \infty) = 1.221$. The expected values of $\gamma$ based on the nonlinear Fokker-Planck equation are 
$\frac{4}{3}=1.333$ and $\frac{40}{33}=1.212$, respectively.} \label{gamma_w}
\end{figure}

\subsection{Sub- and Super-diffusion from the same model}
Finally, with an understanding of the issues involved in taking the continuum limit, we turn to a demonstration of sub- and super-diffusion from the same model. To do so, we have to address two additional issues. The first is that the choice $a=0$ is problematic for sub-diffusion. The reason is that if the distribution is zero for, say, $r>r_{0}$ then with $a=0$ the probability for a walker to jump from any site $r<r_{0}$ to a site $r>r_{0}$ is zero (see Eq.(\ref{zerox})). Even with an initial condition that, technically, has infinite support (such as a Gaussian), the distribution will be numerically zero outside some range so that in the diffusive case the maximum support is limited. To be able to perform arbitrarily long simulations over arbitrarily large lattices, we must have $a>0$ in the jump probabilities (Eq.(\ref{F_G}). The second issue is purely a matter of the choice of problem: we wish to display both sub-diffusion and super-diffusion in the same system while only changing a single control parameter. Because the rates of the two processes are very different, the comparison either necessitates very long runs for the sub-diffusive process or very restricted ranges of exponents for the super-diffusive process. Our compromise presented here was to take somewhat higher elementary probabilities, $p_{n}=10^{-3},n \in [-2,-1,1,2]$, $a=10^{-4}$ and $0.65 \le \alpha \le 2$ and beginning with an initial Gaussian with width $10$ lattice units. The results are shown in Fig.\ref{SubSuper} where, after an initial transient period, the three regimes of diffusive behavior are clearly visible.


\section{Concluding comments}
\label{Comments}
Beginning with a general formalism for describing the statistics of a population of interacting random walkers developed previously \cite{boon-lutsko,lutsko-boon} we have constructed a particular class of interactions that can give rise to sub-diffusion, ordinary diffusion and super-diffusion with the adjustment of a single parameter. This flexibility depends on the introduction of non-local as well as non-linear interactions between the walkers. Not only does the mean-squared displacement show the characteristic behavior of these three regimes, but the systems are diffusive in the strong sense of admitting self-similar solutions. Our model also demonstrates universality: not only does our particular ansatz give the same macroscopic behavior independent of the choice of the parameter $a$, but the sub-diffusive regime reproduces the same nonlinear Fokker-Planck equation as in our previous work that was based on a local interaction \cite{lutsko-boon}. On the other hand, it is worth noting that the self-similar behavior results from self-similar initial conditions: because of the nonlinearity of the generalized diffusion equation, we cannot know whether other initial conditions might evolve into other classes of long time behavior.

We note that our calculations are based on a particular ansatz for the concentration-dependence of the jump probabilities, Eq.(\ref{model_alt}). The ansatz was motivated by simplicity and the requirements on the probabilities such as positivity and normalization. Otherwise, there is nothing special or unique about the choices made here: in particular, the fact that we allow for jumps of length $2$ and the form of the ansatz itself. While we were able to characterize the effect of our model in terms of effective attraction or repulsion of the population of walkers, no attempt was made to relate it to microscopic interactions (e.g. particular force laws). Nevertheless, the underlying physics is simply that an effective repulsion between the walkers gives rise to super-diffusion while an attraction causes sub-diffusion. It is the demonstration that such a mechanism can cause anomalous diffusion that is one of the main physical results of this paper. The actual form of the jump probabilities, given in Eq.(\ref{F_G}), is irrelevant to establishing this point. The determination of simpler or more physical alternatives to Eq.(\ref{F_G}) is a matter for future research.

It is interesting to ask whether the present approach might be related to other mechanisms that generate anomalous diffusion. As discussed in the Introduction, many of these involve non-Markovian elements while here the microscopic random walk has no memory at all: it is simply the interaction between an individual walker and the local concentration of walkers at a particular instant that gives rise to the anomalous behavior. One Markovian mechanism for generating super-diffusion is the L\'{e}vy flight in which the probability to make a jump of a given length decays, asymptotically, as a power of the length. This therefore allows for arbitrarily long jumps in one time-step. However, even though there is a superficial similarity in the occurrence of power-laws in the L\'{e}vy flight and our model, the two mechanisms are completely distinct since here the dynamics does not rely on long-ranged jumps. Indeed, in the numerical example presented above, the jumps were limited to at most two lattice sites and we have obtained similar results allowing for only nearest-neighbor jumps. We do not achieve super-diffusion by making jumps of a given length faster or by allowing jumps to be very long during a fixed time (as in the L\'{e}vy flight) but by a bias in the direction of a jump that is based on the local density of walkers.
So our approach represents a completely independent mechanism for generating anomalous diffusion.

 One aspect of the specific model studied here that deserves discussion is the sensitivity to initial conditions. We found that not all initial conditions led to  self-similar long-time solutions. One reason for this, as discussed above, is the need to respect the various assumptions necessary to derive the generalized-diffusion equation from the microscopic dynamics. However, it is also true that the generalized-diffusion equation is nonlinear so that it is not easy to make general statements about its long time behavior. It is possible that all initial conditions will lead to the same asymptotic state but it is also possible there could be classes of initial conditions that lead to different asymptotic states. It is even possible that the generalized diffusion equation is not ergodic: i.e., that the long-time behavior preserves a memory of the initial condition. Based on numerical experiments beyond those reported here, we do not believe this is the case but a rigorous investigation e.g. based on Kinchin's theorem such as has been made for the fractional diffusion equation (see \cite{Burov}), has yet to be attempted.

One interesting aspect of our results is that the range of scaling exponents that can be generated is restricted to $0 < \gamma \le 1.5$ {(recall that for larger values of $\gamma$ the mean-squared displacement does not exist, see Eq.(\ref{bound}))}. The upper limit is due to the nature of the self-similar solutions which are power-laws. As such, they only give finite result for a finite range of moments of the spatial variable. Normalization (existence of zeroth-order moment) demands that  $\alpha \ge -1$ which is not restrictive in terms of the scaling exponent. However, existence of the mean-squared displacement itself demands that $\alpha > \frac{1}{3}$ thus giving the quoted limit on $\gamma$. In contrast, one typically expects that the upper limit for super-diffusion is $\gamma = 2$ corresponding to ballistic motion. The difference is that ballistic motion is not-stochastic: an initial condition of a delta-function concentration of walkers at the origin would evolve, at best, as two delta-functions moving in opposite directions at constant speed with the only stochasticity occurring in the initial choice of direction. In contrast, the ``free'' limit of our model is the case of no interactions corresponding to ordinary diffusion. Moving away from ordinary diffusion requires turning on interactions and there is no reason to expect that this should ever lead to deterministic, non-stochastic behavior.   
As a matter of fact ballistic motion, although exhibiting mathematically   a form corresponding to $\gamma = 2$, originates from a physical mechanism (such as tracer dispersion \cite{sanchez} in an active medium \cite{dogariu}) different from molecular diffusion where free (ballistic) motion occurs at the microscopic scale only in the mean free path regime. 

Other models, such as the fractional diffusion equation which is based on a continuous time random walk with power-law distributed waiting times, can also give rise to anomalous diffusion. So it is natural to ask the questions:
(i) which approach is appropriate to describe the physics of anomalous diffusion? and (ii) when experimental results are
analyzed, how to discriminate between different approaches in order to establish the underlying mechanism?
 Our view is that all of the proposed mechanisms are potentially relevant and that different mechanisms may be at work in different physical systems. A more pertinent question is therefore how one might distinguish, experimentally, between the different models. Clearly, since they all produce anomalous diffusion in terms of the mean-squared displacement, the test must involve something else. We note, for example, that the probability (or concentration) distribution predicted by the fractional diffusion equation is a stretched-exponential \cite{FFPE_PhysToday, metzler-klafter} which is qualitatively quite different from the power-law distributions predicted by the present model. Therefore, a determination of the distribution would be one powerful method for distinguishing between the microscopic mechanisms. Unfortunately, in practice this is more difficult to extract from experiment than is the mean-squared displacement and so it is seldom available. One notable exception is a recent experiment on morphogen-gradient generation where in fact we have shown that the power-law distributions seem to provide a much better descripton of the data than exponential distributions\cite{boon-lutsko-lutsko}.

\begin{acknowledgments}
The work of JFL was supported by the European Space Agency under contract number
ESA AO-2004-070.
\end{acknowledgments}



\begin{thebibliography}{99}
\bibitem{NoteC}
Note that the advection term in the generalized Fokker-Planck equation reads
${\cal{C}}_{\alpha}\, \frac{\partial }{\partial r}f$ where 
${\cal{C}}_{\alpha}\,=\, \frac{1+a}{2}\,\left({\lambda  }\,f\right)^{\alpha -1}\,M_1$ 
has dimensions $[{\cal{C}}_{\alpha}] = L \, T^{-1}$ and,  in the limit  
$\alpha \rightarrow 1$ and $a=1$, gives the classical drift speed ${\cal{C}}_{\alpha =1}\,=\,M_1$ .
\end{thebibliography}
\end{document}